\input{aipcheck}

\documentclass[
    ,final            
  ]
  {aipproc}

\layoutstyle{6x9}

\begin{document}

\title{Energy functional based on natural orbitals 
and occupancies 
for static properties of nuclei. }

\classification{<21.60.Jz, 21.10.Dr}
\keywords      {Energy Density Functional, Density Matrix Functional Theory}

\author{Denis Lacroix}{
  address={GANIL, CEA and IN2P3, Bo\^ite Postale 5027, 14076 Caen Cedex, France}
}

\begin{abstract}
The possibility to use functionals of occupation numbers and natural orbitals 
for interacting fermions is discussed as an alternative to multi-reference energy 
density functional method. An illustration based on 
the two-level Lipkin model is discussed.  
\end{abstract}

\maketitle


\section{Introduction}

The nuclear many-body problem of $N$ interacting nucleons can be solved exactly only in very 
specific cases or for very small particle numbers. This is due to the large 
number of degrees of freedom involved in such a complex system. Let us for 
instance consider particles interacting through n Hamiltonian written 
as
\begin{eqnarray}
\label{eq:hamil} H =   \sum_{ij} t_{ij} \, a^+_i \, a_j
  + \frac{1}{4} \sum_{ijkl} \tilde{v}_{ijkl} \, a^+_i \, a^+_j \, a_l \, a_k  + \cdots
\end{eqnarray}   
Then the exact ground state energy can be written as 
\begin{eqnarray}
E_{\rm Exact}(\gamma^{(1)}, \gamma^{(2)}, ...) =  \sum_{ij} t_{ij} \, \gamma^{(1)}_{ji} + \frac{1}{4} \sum_{ijkl} \tilde{v}_{ijkl} \, \gamma^{(2)}_{kl , \, ij} + \cdots,
\label{eq:funccomplex}
\end{eqnarray} 
where $\gamma^{(1)}_{ji} \equiv \langle a^+_i \, a_j  \rangle$, $\gamma^{(2)}_{kl, \, ij} \equiv \langle  a^+_i \, a^+_j \, a_l \, a_k  \rangle$, ... denote the one-, two-, ... body density matrices
that contain all the information on the one-, two-...body degrees of freedom respectively. A natural way 
to reduce the complexity of this problem is to assume that 
at a given level, the $k-$body (and higher-order) density matrices becomes a functional of the lower-order ones. This is what is done for 
instance in the Hartree-Fock (HF) approximation where all $k$-body density matrices (with $k \ge 2$) become a functional of $\gamma^{(1)}$. Unfortunately, the HF theory applied to the nuclear many-body problem in terms of the 
vacuum Hamiltonian 
is a poor approximation and Many-Body theories beyond HF are necessary.        

The introduction of Energy Density Functional (EDF) approaches in the 70's was a major breakthrough (see for instance \cite{Ben03}
for a recent review). 
In its simplest form, the EDF formalism  
starts with an energy postulated as a functional of $\gamma^{(1)}$, 
the latter being built out of a Slater Determinant. 
Then the ground state energy is obtained by minimizing the energy with respect
to $\gamma^{(1)}$, i.e. 
\begin{eqnarray}
E_{\rm Exact} &\simeq& {\cal E}_{\rm MF}(\gamma^{(1)}) 
\label{eq:simpleedf}
\end{eqnarray}
Parameters are generally adjusted on specific experimental observations and therefore encompass directly many-body correlations.
Current EDF uses a generalization of eq. (\ref{eq:simpleedf}) obtained by considering quasi-particle vacua as trial states. By making 
explicit use of symmetry breaking, such a functional called hereafter Single-Reference (SR-) EDF is able to account for static 
correlation associated with pairing and deformation. Actual SR-EDF takes the form
\footnote{Note that the denomination "mean-field" or the
separation into a "mean-field" like 
and "correlation" like is completely arbitrary since, as we mention previously, the so-called "mean-field" part already 
contains correlation much beyond a pure Hartree-Fock approach.}:
\begin{eqnarray}
E_{\rm Exact} &\simeq& {\cal E}_{\rm MF}(\rm \gamma^{(1)}) + {\cal E}_{\rm Cor}(\kappa . \kappa^*)  \label{eq:epair}
\end{eqnarray} 
where $\kappa$ denotes the anomalous density. To restore symmetries and/or incorporate dynamical correlations, 
guided by the Generator Coordinate Method (GCM), a second level of EDF implementation, namely Multi-Reference (MR-) EDF 
is introduced. Recently, difficulties with the formulation and implementation of 
have been encountered in MR-EDF. A minimal solution has been proposed in ref. \cite{Lac09b,Ben09,Dug09}.  Besides these 
problems, the authors of ref. \cite{Lac09b} have pointed out the absence of a rigorous theoretical framework for the MR EDF approach. 
At the heart of the problem is the possibility to break symmetries in functional theories and then restore 
them using configuration mixing. This issue needs to be thoroughly addressed in the future.

In this context, it is interesting to see if extensions of the functional used at the SR-EDF level can grasp part of the effects that for standard functionals require the MR level. It is worth realizing that, in the canonical basis for which 
$\gamma^{(1)} = \sum_i | \varphi_i \rangle n_i \langle \varphi_i |$, we have  
\begin{eqnarray}
{\cal E}_{\rm Cor} (\kappa.\kappa^*) &=& {\cal E}_{\rm Cor}[\{\varphi_i, n_i \} ]  = \frac{1}{4} \sum_{i,j} \bar v^{\kappa \kappa}_{i\bar i j \bar j} \sqrt{n_i (1-n_i)} \sqrt{n_j (1-n_j)}, \label{}
\end{eqnarray}
and therefore, the energy can be regarded as a functional of natural orbitals $\varphi_i$  and occupation numbers $n_i$. 
As a matter of fact, for electronic systems, Gilbert has generalized the Kohn-Sham theory and 
shown that the exact energy of a system can be obtained by minimizing 
such a functional \cite{Gil75} leading to the so-called Density Matrix Functional Theory (DMFT). 
The possibility to consider occupation numbers as building blocks of the nuclear energy 
functional has recently been discussed in ref. \cite{Pap07,Ber08}. Two levels of theory can be developed along the line of Gilbert's idea 
(i) either, functionals in the strict Gilbert framework can be designed. In that case, since the density 
identify with the exact density at the minimum, it should respect all symmetries of the bare Hamiltonian. 
(ii) or we exploit the concept of symmetry breaking. In the latter case, similarly to the SR-EDF, strictly speaking
we cannot anymore rely on the theorem, but we may gain better physical insight with relatively simple functionals. 

   
\section{Application to the Lipkin model and discussion}
\begin{figure}[t!]
\includegraphics[height=7.cm]{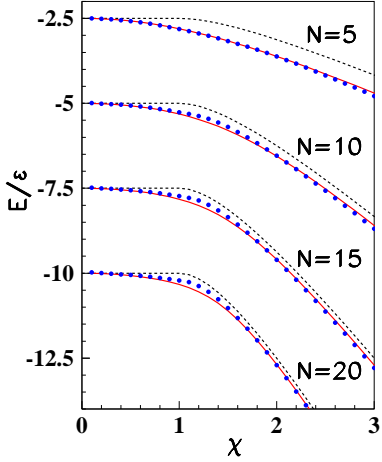}
\caption{Exact ground state energy (solid lines) displayed 
as a function of $\chi$ for $N=5$ to $20$ resp. from top to bottom.
In each case, the corresponding HF (dashed line) and DMFT (filled circle) minimum energy are shown.
The DMFT calculation is performed using the mean-field and correlation energy resp. given by 
Eq. (\ref{eq:emflipkin}) and Eq. (\ref{eq:coreta}) with $\eta(N) =1.5 ~{N}^{-2/3}$ (Adapted from \cite{Lac09a}).} 
\label{fig:chi}
\end{figure}

The descriptive power of DMFT  is illustrated here in the two-level Lipkin 
model \cite{Lip65}. In this model, the Hartree-Fock (HF) theory fails 
to reproduce the ground state energy whereas configuration mixing like Generator Coordinate Method (GCM) provides
a suitable tool \cite{Rin80,Sev06}. Therefore, the two-level Lipkin model is perfectly suited both to illustrate that 
DMFT could be a valuable tool and to provide an example of a functional for system with a 
"shape" like phase-transition. In this model, one considers $N$ particles distributed in two N-fold degenerated shells separated 
by an energy $\varepsilon$. The associated Hamiltonian is given by $H = \varepsilon J_0 - \frac{V}{2} (J_+ J_+ + J_- J_-)$
where $V$ denotes the interaction strength while $J_0$, $J_\pm$ are the quasi-spin operators defined as 
$J_0 = \frac{1}{2} \sum_{p=1}^{N} (c^\dagger_{+,p}c_{+,p} - c^\dagger_{-,p}c_{-,p})$, $J_+ = \sum_{p=1}^{N} c^\dagger_{+,p}c_{-,p}$
and $J_- = J_+^\dagger$.
$c^\dagger_{+,p}$ and $c^\dagger_{-,p}$ are creation operators associated with the upper and lower levels respectively. 
Due to the specific form of the Lipkin Hamiltonian , $\gamma^{(1)}$ simply writes in the natural basis as
$\gamma^{(1)} = \sum_{p=1}^N \Big\{| \varphi_{0,p} \rangle n_0 \langle \varphi_{0,p} | + 
| \varphi_{1,p} \rangle n_1 \langle \varphi_{1,p} | \Big\}$
with $n_1 = (1-n_0)$. Introducing the angle $\alpha$ between the state $| - , p \rangle$ and $| \varphi_0 , p \rangle$, leads to the following 
mean-field functional \cite{Lac09a}
\begin{eqnarray}
{\cal E}_{\rm MF}(\{ \varphi_{i,p}, n_i \}) =  {\cal E}_{\rm MF}(\alpha,n_0) = -\frac{\varepsilon}{2} N  \Big\{ \cos(2\alpha) (2n_0 -1) 
+ \frac{\chi}{2} \sin^2(2\alpha) (2n_0 -1)^2 \Big\} .
\label{eq:emflipkin}
\end{eqnarray} 
where $\chi = V(N-1) / \varepsilon$. This expression is easily obtained by generalizing the Hartree-Fock case (recovered here if $n_0=1$). The main challenge 
of the method is to obtain an accurate expression for ${\cal E}_{\rm Cor}$. To get the functional, clearly identified cases 
from which properties of the functional could be inferred have been used\cite{Lac09a}, namely the $N=2$ case and the large $N$ limit.  
In the two-particles case, the correlation energy can be analytically obtained and reads
\begin{eqnarray}
{\cal E}^{^{{N=2}}}_{\rm Cor}(\alpha, n_0) &=& - 2 V \Big\{\sin^2(2\alpha) n_0(1-n_0) 
+\left(\sin^4(\alpha) + \cos^4(\alpha)\right) \sqrt{n_0 (1-n_0)} \Big\}.
\end{eqnarray} 
A simple extension of the $N=2$ case for larger number of particles is to assume that each pair contributes independently from the others 
leading to ${\cal E}^{N}_{\rm Cor} = \left[N(N-1)/2\right] {\cal E}^{^{{N=2}}}_{\rm Cor}$. However, such a simple 
assumption leads to a wrong scaling behavior in the large $N$ limit. Indeed, in this case, 
${\cal E}^{N}_{\rm Cor} \propto N^2$ as $N$ tends to infinity while a $N^{4/3}$ scaling is expected 
\cite{Dus04}. To obtain the correct limit, a semi-empirical factor $\eta(N)$ can be introduced such that
\begin{eqnarray}
{\cal E}^{^{{N\ge 3}}}_{\rm Cor}(\alpha , n_0) =  \eta(N) \frac{N(N-1)}{2} {\cal E}^{^{{N=2}}}_{\rm Cor}(\alpha , n_0) ,
\label{eq:coreta}
\end{eqnarray}
with $\eta(N) = c N^{-2/3}$. The value $c=1.5$ has been retained using a fitting procedure. Examples of results obtained by minimizing the 
functional given by Eqs. (\ref{eq:emflipkin}) and (\ref{eq:coreta}) are shown in Fig. \ref{fig:chi} for different 
particle numbers and interaction strengths. In all cases, a very good agreement, much better than the HF case is found.    

The Lipkin example suggests that DMFT can be a valuable tool for 
describing ground state of a many-body system when symmetry breaking plays
a significant role. The functional designed here is exact only in the $N=2$.
Note that the functional proposed here breaks signature symmetry and therefore 
enters into the level (ii) of functional discussed in the introduction.
The Lipkin model is however rather schematic and cannot be used 
as a guidance for realistic situations. The possibility to design a new accurate
functional for nuclei remains a challenging problem.


\begin{theacknowledgments}
The author thanks M. Assi\'e, B. Avez, M. Bender, T. Duguet,
C. Simenel, O. Sorlin and P. Van Isacker for enlightening discussions
at different stages of this work and T. Papenbrock for useful
remarks on the scaling behavior in the Lipkin model.
\end{theacknowledgments}




\end{document}